\documentstyle[12pt,emlines2]{article}
\begin{document}
\large
\begin{sloppypar}

\title{\bf\it ASYMMETRY PARAMETER ROLE IN DESCRIPTION \\ OF PHASE STRUCTURE
OF LATTICE GLUODYNAMICS \\ AT FINITE TEMPERATURE.}

\author{ L.~A.~Averchenkova,\and K.~V.~Petrov,\and V.~K.~Petrov,\and 
G.~M.~Zinovjev \\
\it Bogolyubov Institute for Theoretical Physics\\
National Academy of Sciences of Ukraine\\
Kiev 143, UKRAINE \thanks{e-mail: hedpitp@gluk.apc.org, 
gezin@hrz.uni-bielefeld.de, lily@knb.kiev.ua}}

\maketitle

\begin{abstract}
The role of lattice asymmetry parameter $\xi$ in description of the $SU(2)$
gluodynamics phase structure at finite temperature is studied analytically.
The fact that renormalization group relations which permit to remove the
lattice asymmetry parameter from the thermodynamical quantities in
the "naive" limit don't do the same in the approximation $SU(N)\simeq Z(N)$
keeping the "naive" limit results at the same time was point out.
An additional condition which would fix $\xi$ is needed for this dependence
removal.
\end{abstract}

\begin{center}
INTRODUCTION.
\end{center}

Following the heart point of the renormalization group theory
after any regularization removal (in particular, lattice one)
observed quantities should go to their physical values and should not
depend on the renormalization scheme. So, although
going to the continuum from a symmetrical
lattice and from asymmetrical one are two different
renormalization procedures in the gauge fields theory
it is natural to expect that in the
continuum limit no lattice asymmetry effects will be revealed.
At the same time, strictly speaking,
the renormalization group trajectories are universal
(i.e., they don't depend on the renormalization scheme)
only near the critical point at zero coupling
while at intermediate and strong couplings
this statement is still questionable. It is quite probable
the trajectories will depend on the degree of lattice asymmetry \cite{shigem}.
Moreover, today's level of this problem understanding does not permit to
exclude other possible revealings of the lattice asymmetry.

In this paper we consider the finite temperature $SU(N)$-gluodynamics
on an asymmetrical lattice

\begin{math}
a_{\mu}\neq a_{\nu};\;\; \mu=0,1,2,3;\;\; a_{0}N_{0}=\beta;\;\;
a_{k}N_{k}=L_{k};\;  k=1,2,3.
\end{math}\\
$a_\mu$ -- the lattice spacing along the direction $\mu$.
The most wide-spread example of such asymmetry introducing is the Hamiltonian
limit \cite{shigem,hasenf}
(where the lattice distance along the space directions is fixed and
the spacing along the time direction is going to zero).
The renormalization procedure independence actually means
that renormalized couplings of both theories (for example,
Hamiltonian and Euclidean $SU(N)$ lattice gauge theories) should be get equal.
This naturally imposes the relation on the scales
$\frac{\Lambda_{E}}{\Lambda_{H}}$.
Following chosen approach \cite{shigem, hasenf}
this is enough for two renormalized theories to give the "same physics".
Indeed, if in the standard Wilson's action
\begin{equation}
-{\cal S} = \kappa_{\mu \nu} \sum_{x,\mu < \nu} {\rm Re}\: {\rm Tr} (1-\frac{1}{N_c}
U_{\mu}(x)U_{\nu}(x+\mu)U_{\mu}^{+}(x+\nu)U_{\nu}^{+}(x))
\label{1}
\end{equation}
we restrict ourselves to only "naive" limit (which assumes the expansion
of $SU(N)$ matrices around I-matrix, close to I-matrix or any fixed matrix)
\begin{equation}
U_{\mu}(x) = e^{ia_{\mu}g_{\mu}{\cal A}_{\mu}(x)} \simeq 1+ia_{\mu}g_{\mu}
{\cal A}_{\mu}(x)
\label{2}
\end{equation}
and we take
\begin{equation}
\kappa_{\mu \nu}=\frac{2N}{g_{\mu}g_{\nu}}
\frac{a_{0}a_{1}a_{2}a_{3}}{a_{\mu}^2 a_{\nu}^2}
\label{3}
\end{equation}
then at least in the Hamiltonian (H) and the Euclidean (E)
[$a_{\nu}=a;$ $g_{\nu}^2 \rightarrow g_{E}^2$]
limits the spectra of hamiltonians coincide.
Moreover, not only in the classical case ($g_{E}^2 = g_{H}^2$), but in
the quantum case ($g_{E}^2 \simeq g_{H}^2 + O(g_{H}^4)$) too.

However, investigation of the finite temperature $SU(N)$ lattice gauge theory in
"naive" limit is carried out using $SU(N)$ matrices expansion around fixed
matrix thereby leaving the center $Z(N)$ out of the game.
But the role of the gauge group center in the finite temperature QCD phase
structure is well-known. Therefore, we will study the lattice asymmetry
effects contributed by the center $Z(N)$. To do this we will use the approach
$SU(N)\simeq Z(N)$, i.e., we will consider the finite temperature lattice
gauge theory in which
the link variables are the center elements of the $SU(N)$ gauge group
\begin{equation}
\sigma_{\mu}(x)=e^{\frac{2\pi iq_{\mu}}{N}}; \;q_{\mu}=0,1,2 \ldots  N-1
\label{4}
\end{equation}

On the other hand,
for theories with gauge groups which have a trivial center (for example,
$SU(N)/Z(N)$) it may be quite reasonable to restrict ourselves to "naive" limit,
whereas for the $SU(N)$ gauge theories such a restriction
requires special substantiation because there is no naive limit for the center Z(N)
in common sense.
Building the continuum quantum field theory becomes possible
only if $Z(N)$ lattice gauge system undergoes the second
order phase transition at zero of the renormalization group function
\cite{creutz, brez, pena}.

To ensure agreement with the "naive" limit results
we chose $\kappa_{\mu\nu}$ ($\ref{3}$) to
guarantee the coincidence of naive limits for the models with $SU(N)$ gauge
group on the symmetrical lattice and asymmetrical one.

In the future we will calculate the corrections to our approximation.
As a justification of our approach we should note that
effect of the quantum fluctuations, which are taken into account
besides the $Z(N)$ topological excitations,
amounts to simple change in the coupling constant
\cite{hasenf,yon,lusch}
\begin{equation}
\kappa_{new}=\kappa_{old}- (N^2-1)/4
\label{5}
\end{equation}
Additional term (\ref{5}) depends neither on $a_\mu$ nor on $\kappa_\mu$.
Therefore, it does not depend on the lattice asymmetry.
So, there is a reason to hope that
including the quantum fluctuations besides of $Z(N)$ solutions
in our theory on asymmetrical lattice
will not change (\ref{5}), i.e.
will just result in the finite renormalization of couplings.

It will be enough for our purposes to consider the simplest case
of lattice asymmetry, when
$\kappa_{0n}\equiv\kappa_{\tau}$ and
$\kappa_{nm}\equiv\kappa_{\sigma}$,
i.e., we allow the couplings to be
different for plaquettes lying in the temporal and in the spatial planes.
Our results may be easily generalized to the case (\ref{3}).
We will also limit ourselves to the analyses of four limiting cases:
\begin{equation}
1)\;\kappa_{\sigma}\rightarrow 0;\;\;\;
2)\;\kappa_{\tau}\rightarrow 0;\;\;\;
3)\;\kappa_{\sigma}\rightarrow \infty;\;\;\;
4)\;\kappa_{\tau}\rightarrow \infty;\;\;\;
\label{6}
\end{equation}
(the estimations of phase structure in the whole plane
$\kappa_\tau\otimes\kappa_\sigma$ we are planning to present in next paper).

As known our Wilson's action is splitting into two parts:
\begin{eqnarray}
-{\cal S}(\sigma) = \kappa_{\sigma} \sum_{P_{\sigma}}
\left(1-\frac{1}{N}{\rm Re}\:{\rm Tr}\sigma_{n}(\vec{x})\sigma_{m}(\vec{x}+n)
\sigma_{n}^{+}(\vec{x}+m)\sigma_{m}^{+}(\vec{x})\right) + \nonumber \\
\kappa_{\tau} \sum_{P_{\tau}}\left(1 - \frac{1}{N}{\rm Re}\:{\rm Tr}
\sigma_{0}(\vec{x},\tau)\sigma_{n}(\vec{x},\tau+0)
\sigma_{0}^{+}(\vec{x}+n,\tau)\sigma_{n}^{+}(\vec{x},\tau)\right)
\label{7}
\end{eqnarray}
where  $\kappa_{\sigma}=2N a_{\tau}/{g_{\sigma}^2 a_{\sigma}}$
and $\kappa_{\tau}=2N a_{\sigma}/{g_{\tau}^2 a_{\tau}}; \;\;
a_{\sigma}(a_{\tau})$ is the spatial (temporal) spacing and the sum
$P_{\sigma}(P_{\tau})$ runs over
all purely space-like (time-like) plaquettes.

Let's first consider  $\gamma_\sigma\equiv{\rm th}\kappa_\sigma$ and
$\gamma_\tau\equiv{\rm th}\kappa_\tau$ independent temporary ignoring the
dependence of $g_\sigma(g_\tau)$ on $a_\sigma(a_\tau)$ through the
renormalization group relations, i.e., supposing that by varying $a_\sigma$
and $a_\tau$ we can get to any point of the square
($0\leq\gamma_\tau\leq 1;\;\;0\leq\gamma_\sigma\leq 1$).
We will investigate each of the limiting cases separately,
joining the results for the couplings critical values
through the same parameter.

\vspace{1.5cm}
\begin{center}
{\it THE AREA OF SMALL $\kappa_{\sigma}$.}
\end{center}

In the case $\gamma_{\sigma}\simeq 0$
we may throw off the magnetic part of action
thereby leaving only the spin configurations with time-like plaquettes.
In the static gauge we can sum over the space spin configurations
$\{\sigma_{n}\}$,
finally gaining the Ising model which is much known of.

Now let us dwell upon that.
Fixing the static gauge
$\sigma_{0}(\vec{x},\tau)=\omega_{\vec{x}}$
we consider the following gauge transitions
\begin{eqnarray}
\sigma_{n}(\vec{x},\tau) \rightarrow \omega_{\vec{x}}^{\tau}
\sigma_{n}(\vec{x},\tau)\omega_{\vec{x}+n}^{-\tau} \nonumber \\
\sigma_{n}^{+}(\vec{x},\tau) \rightarrow \omega_{\vec{x}+n}^{-\tau}
\sigma_{n}^{+}(\vec{x},\tau)\omega_{\vec{x}}^{\tau}
\label{8}
\end{eqnarray}

Imposing the periodical boundary conditions in temporal direction
upon $\sigma$-matrices:
$\sigma_{n}(\vec{x},\tau=1)\equiv\sigma_{n}(\vec{x},\tau=N_{\tau})$ ,
all matrices
$\omega_{\vec{x}}$
are grouped on the last links
resulting in the Polyakov's loops
\begin{equation}
\Omega_{\vec{x}}=\prod_{\tau=1}^{N_\tau} \omega_{\vec{x}}
=\omega_{\vec{x}}^{N_\tau}
\label{9}
\end{equation}
We come to the following action
\begin{eqnarray}
-{\cal S}_{P_{\tau}} = \kappa_{\tau}\sum_{\vec{x},n}
\left[
   \sum_{\tau=1}^{N_{\tau}-2}
   \begin{array}{c}
      \left(
         1-\frac{1}{N}{\rm Re}\:{\rm Tr}
         \sigma_{n}(\vec{x},\tau)\sigma_{n}^{+}(\vec{x},\tau+1)
      \right)
   \end{array}
\right. + \nonumber \\
\left.
   \begin{array}{c}
      \left(
         1-\frac{1}{N}{\rm Re}\:{\rm Tr}
         \Omega_{\vec{x}}      \sigma_{n}    (\vec{x},N_{\tau}-1)
         \Omega_{\vec{x}+n}^{+}\sigma_{n}^{+}(\vec{x},1)
      \right)
   \end{array}
\right]
\label{10}
\end{eqnarray}
Summing over all configurations
$\{\sigma_{n}\}$ in $Z(2)$ gauge system
we get
\begin{eqnarray}
-{\cal S}_{P_{\tau}}=\tilde{\kappa_{\tau}} \sum_{\vec{x},n}
\Omega_{\vec{x}}\Omega_{\vec{x}+n}^{\star}  \\ \label{f11}
{\rm th}\tilde{\kappa_{\tau}}=({\rm th}\kappa_{\tau})^{N_{\tau}}=\tilde
{\gamma_{\tau}}\nonumber
\end{eqnarray}

As known, for the Ising model there exists the critical value of the coupling
$\tilde{\gamma_{\tau}}=\gamma_c$
separating two phases.
Consequently, the partition function (\ref{10}) will have
a critical point at
$\gamma_{\tau}^c = \gamma_{c}^{1/N_{\tau}}$.

\vspace{1.5cm}
\begin{center}
{\it THE AREA OF SMALL $\kappa_{\tau}$.}
\end{center}

In the other limiting case
$\gamma_{\tau} \simeq 0$
the functional integral is highly peaked about
configurations with space-like plaquettes
and we may throw off the electric part of action.
In this case the partition function turns out to fall to
$N_{\tau}$ equivalent disconnected contributions
each of which is from a separate 3-dimensional layer.
We get the set of standard 3-dimensional Wegner models
\begin{equation}
{- S_{P_{\sigma}}} \simeq \sum_{\tau=1}^{N_{\tau}}\sum_{x,n,m}
\kappa_{\sigma}\sigma_{n}\sigma_{m}
\sigma_{n}^{\star}\sigma_{m}^{\star}
\label{12}
\end{equation}
There is no interaction between the layers, so summing over
$\sigma_n(x,\tau)$
may be done independently for every $\tau=const$ layer.

The lattice dual to original hypercubical lattice
can be constructed by
shifting the lattice by half a lattice spacing in each direction
(see, for example, \cite{savit}).
Geometrical duality transforms $q$-dimensional manifolds
into $(d-q)$-dimensional ones.
"Dual coupling constant"
$\tilde{\kappa}$
for the $Z(2)$ gauge theory
$\tilde{\kappa_{\sigma}}=-\frac{1}{2}\ln{\rm th}\kappa_{\sigma}$
is a monotonically decreasing function of the "original" coupling.
The set of 3-dimensional Wegner models
transforms into the set of
3-dimensional Ising models with spins in sites
under duality transformations.
These Ising models exhibit the transition at
$\gamma_c$ simultaneously.
Therefore, for the partition function (\ref{12}) there is a critical point at
$\gamma_{\sigma}^c=\frac{1-\gamma_c}{1+\gamma_c}$.

\vspace{1.5cm}
\begin{center}
{\it TWO ANOTHER LIMITING CASES: $\kappa_\tau\rightarrow\infty$ AND
$\kappa_\sigma\rightarrow\infty$.}
\end{center}

When considering the duality transformation in 4-di\-men\-sio\-nal space-time
it should be pointed out that just space-like plaquettes
transform into time-like ones.
In the other words,

\begin{equation}
\kappa_{\sigma}^{\prime}=-\frac{1}{2}\ln{\rm th}\kappa_{\tau}\;\;\; or \;\;\;
\gamma_{\sigma}^{\prime}=\frac{1-\gamma_{\tau}}{1+\gamma_{\tau}}
\label{13}
\end{equation}
and vice versa,
$\kappa_{\tau}^{\prime}=-\frac{1}{2}\ln{\rm th}\kappa_{\sigma}$.

This statement becomes clear from the following.
Let us rewrite the partition function of $Z(2)$ system in the form
\begin{eqnarray}
{\cal Z}
&=&\sum_{\{\sigma\}}e^{\kappa_{\mu\nu}\sum_{x,\mu\nu}\Box_{x,\mu\nu}}\\ \nonumber
&=&\sum_{\{\sigma\}}\prod_{x,\mu\nu}{\rm ch}\kappa_{\mu\nu}
    (1+\Box_{x,\mu\nu}{\rm th}\kappa_{\mu\nu}) \\ \nonumber
&=&e^{-Nf} \sum_{\{\sigma\}} \prod_{x,\mu\nu} \sum_{q=\pm 1}
      (\Box_{x,\mu\nu}{\rm th}\kappa_{\mu\nu})^{\frac{q_{x,\mu\nu}+1}{2}} \\
&=&e^{-Nf} \sum_{\{q\}} \prod_{x,\mu\nu}
      (e^{\ln{\rm th}\kappa_{\mu\nu}})^{\frac{q_{x,\mu\nu}+1}{2}}
         \prod_{links} \sum_{\sigma=\pm 1}
            (\sigma)^{\sum_{\mu=-3,\mu\neq\pm\nu}^{3}
            \frac{q_{x,\mu\nu}+1}{2}} \nonumber \\
&=& e^{-Nf}\sum_{\{q\}} \prod_{x,\mu\nu}
      (e^{\ln{\rm th}\kappa_{\mu\nu}})^{\frac{q_{x,\mu\nu}+1}{2}}
         \prod_{links}2\delta_{2}
             (\sum_{\mu=-3,\mu\neq\pm\nu}^{3}
                 \frac{q_{x,\mu\nu}+1}{2}
             )\nonumber
\label{14}
\end{eqnarray}
where $\Box_{x,\mu\nu}=\sigma_\mu(x)\sigma_\nu(x+\mu)\sigma_{\mu}^{\star}
(x+\nu)\sigma_\nu^{\star}(x)$ and \\ $f=N_\tau N_\sigma^3 \sum_{\mu\nu}
\ln{\rm ch}\kappa_{\mu\nu}$.

We introduced a new set of variables $\{q\}$ - one for each plaquette.
The partition function is not equal to zero only if
$q_{x,\mu\nu}$ satisfies the following condition
on the sum over six $q_{x,\mu\nu}$
(associated with six plaquettes which adjoin the link $x,\nu$, see fig.1):
\begin{equation}
\frac{1}{2}\sum_{\mu=-3;\mu\neq\pm\nu}^{3}(q_{x,\mu\nu}+1)=0_{\bmod 2} \;\;\; or
\; \sum_{\mu=-3;\mu\neq\pm\nu}^{3}q_{x,\mu\nu}=2_{\bmod4}
\label{15}
\end{equation}
The solution of last equation can be found
if we associate every $q_{x,\mu\nu}$
with one of the cube plane and
\begin{eqnarray}
q_{x,\mu\nu}
&=& s_\rho(x)s_\lambda(x+\rho)s_\rho^\star(x+\lambda)s_\lambda^\star
(x)\nonumber\\
\nu &\neq& \mu\neq\rho\neq\lambda
\label{16}
\end{eqnarray}
where the dual link variable $s_\rho(x)$ is the element of the $Z(2)$ group.
It becomes intuitively evident, if we consider the starting case
when all $s$ are equal 1.
It dictates for $\sum_{\mu\neq\pm\nu}q_{x,\mu\nu}$ to be equal
$6_{\bmod 4}=2_{\bmod 4}$.
Every link enters the solution twice
(because the plaquettes form a cube)
and changing the sign of a link to opposite
results in changing $\sum_{\mu=-3;\mu\neq\pm\nu}^{3}q_{x,\mu\nu}$
only by $\pm 4$.

Consequently, in the plane $\kappa_{\tau}\otimes\kappa_{\sigma}$
there is the self-\-du\-a\-li\-ty line (see fig.2).
R. Balian, J.M. Drouffe, C. Itzyk\-son \cite{balian} pointed out
the possibility of
the critical behaviour at $\kappa_c=0.44$
for the 4-di\-men\-sio\-nal $Z(2)$ pure gauge theory on symmetrical lattice
supposing this critical point is single.

So, now we realize that under duality transformation our
original 4-dimensional theory (\ref{7}) transforms into the same one
but with new coupling constants.
\begin{equation}
-{\cal S}^{\prime}=\kappa_{\sigma}^{\prime}\Box_{\sigma}+
\kappa_{\tau}^{\prime}\Box_{\tau}
\label{17}
\end{equation}

For this dual representation we can consider the two limiting cases
in precisely the same way we have done for the original theory.

The case $\gamma_{\tau}^{\prime}={\rm th}\kappa_{\tau}^{\prime}\simeq 0$
for the original theory means $\gamma_{\sigma}\simeq 1$
in according to ($\ref{13}$).
Throwing off time-like part of the action
we can go again from the set of 3-dimensional Wegner models
to the set of standard 3-dimensional Ising models
via the duality transformation.
And after that we may come back to variables of the original theory.
\begin{eqnarray}
\kappa_{\sigma}^{\prime}\sum_{P_{\sigma}}\Box_{\sigma}\longrightarrow
\tilde{\kappa_{\sigma}^\prime}\sum_{\vec{x}n}  s_{\vec{x}} s_{\vec{x}+n}^\star
\nonumber \\
{\rm th}\tilde{\kappa_{\sigma}^\prime}=\tilde{\gamma_{\sigma}^\prime}=
\frac{1-\gamma_{\sigma}^{\prime}}{1+\gamma_{\sigma}^{\prime}} \\
\gamma_{\sigma}^{\prime c}=\frac{1-\gamma_c}{1+\gamma_c};\;\;\;
\gamma_{\tau}^{c}=\frac{1-\gamma_{\sigma}^{\prime c}}{1+\gamma
_{\sigma}^{\prime c}}=\gamma_c\nonumber
\label{18}
\end{eqnarray}
Here tilde $\tilde{\kappa}$ (prime $\kappa^\prime$) means
duality transformations in three (four) dimensions, respectively.
The original theory undergoes the phase transition at
$\gamma_{\sigma}\simeq 1$ and $\gamma_{\tau}=\gamma_c$.

Similarly, let's consider the limit $\gamma_{\sigma}^{\prime}\simeq 0$
(it corresponds $\gamma_{\tau}\simeq 1$ for the original theory)
as we have done already for the limit $\gamma_{\sigma}\simeq 0$.
We can find the phase transition at
$\gamma_{\tau}\simeq 1;\;\;\gamma_{\sigma}=\frac{1-\gamma_{c}^{1/N_{\tau}}}
{1+\gamma_{c}^{1/N_{\tau}}}$.
We may depict all these critical points in the plane
$\gamma_{\tau}\otimes\gamma_{\sigma}$ (fig.3).

To avoid arising infinite constant
in the partition function at $\gamma_\sigma=1$ and at $\gamma_\tau=1$,
we took $\gamma_\sigma$ and $\gamma_\tau$
close to 1 but not equal 1 exactly.

\vspace{1.5cm}
\begin{center}
{\it ANOTHER APPROACH TO THE CONSIDERATION OF LIMITS: $\kappa_{\tau}
\rightarrow\infty$ AND $\kappa_{\sigma}\rightarrow\infty$.}
\end{center}

There is also another method to find critical points 3 and 4 (fig.3),
different  from the above duality transformation approach.
Let us build up the effective action
for the case $\kappa_{\sigma}\rightarrow\infty$.
It is obvious that space-like plaquette variables in this case
$\Box_{nm}\equiv\Box_\sigma=1$.
The gauge is completely fixed if we have a maximal tree,
i.e. tree which will have a closed loop after adding one more link to it.

We may choose the maximal tree in two dimensions like a snail
as shown at the fig.4.
By analogy we may obtain such a maximal tree for 3 dimensions.
That choice of the maximal tree provides that the equation
$\Box_{\sigma}=1$ will have only single solution: $\sigma_{x}^n =1$.
Really, if we consider the first plaquette,
which have three link variables equal 1 due to the gauge condition
and provided with $\Box_{\sigma}=1$,
the fourth link is to be equal 1 also.
Now we can apply these considerations to the second plaquette in the snail
and so on, getting finally $\sigma_{x}^n =1$.
As a result, only temporal links will survive in the action.
\begin{equation}
-{\cal S}=\gamma_{\tau}\sum_{\tau}\sum_{\vec{x}n}
\sigma_{0}(\vec{x},\tau)\sigma_{0}(\vec{x}+n,\tau)
\label{19}
\end{equation}

We've got the set of 3-dimensional independent Isings with
critical point at
$\gamma_{\sigma}\simeq 1;\;\;\;\gamma_{\tau}=\gamma_{c}$
for our original system.

In the limit $\gamma_{\tau}\simeq 1$
the plaquette variables $\Box_{0n}\equiv\Box_{\tau}=1$.
In the Hamiltonian gauge
($\sigma_{0}=1)\;\;\;\;\sigma_{n}(t)=\sigma_{n}(t+1)=\bar{\sigma}_{n}$
are time independent.
We come to the static 3-dimensional Wegner system
\begin{equation}
-{\cal S}=N_{\tau}\kappa_{\sigma}\sum_{\vec{x}n}\bar{\sigma}_{n}(\vec{x})
\bar{\sigma}_{m}(\vec{x}+n)\bar{\sigma}_{n}(\vec{x}+n+m)
\bar{\sigma}_{m}(\vec{x}+m)
\label{20}
\end{equation}
which transforms into the Ising model under duality transformations.
\begin{eqnarray}
N_{\tau}\kappa_{\sigma}^{c} &=& -\frac{1}{2}\ln\gamma_{\sigma}^{\prime c}\nonumber \\
\kappa_{\sigma}^{c} &=& -\frac{1}{2}\ln\gamma_{\sigma}^{\prime c 1/N_{\tau}} \\
\gamma_{\sigma}^{c} &=& \frac{1-\gamma_{c}^{1/N_{\tau}}}
{1+\gamma_{c}^{1/N_{\tau}}}\nonumber
\label{21}
\end{eqnarray}
So, this alternative method confirms the results obtained previously.

\vspace{1.5cm}
\begin{center}
{\it THE POTENTIAL BETWEEN PROBE SOURCES.}
\end{center}

We would like to estimate the connected correlation function
$\langle\sigma_{x}\sigma_{x+R}\rangle$
for the Ising model with different couplings in each direction
within the spherical model.
The crucial point is the following condition:
\begin{equation}
\frac{1}{N_\sigma^3}\sum_{x}\sigma_{x}^{2}=1
\label{22}
\end{equation}
Then
\begin{equation}
{\cal Z}
= \sum_{\{\sigma\}} \int_{{\it\bf c}-i\infty}^{{\it\bf c}+\infty}
   \frac{d\alpha}{2\pi i}
   e^{\alpha N_\sigma^3 - \alpha \sum_x \sigma_x^2
       + \frac{1}{2} \sum_{x,n} \kappa_n^\prime
      \sigma_x \sigma_{x+n}}
\label{23}
\end{equation}
where
$\kappa_{n}^\prime=\kappa_{mk}$
the constant ${\it\bf c}$ is chosen to ensure
the legitimacy of interchanging the integration and summation order.
It means that ${\it\bf c}$ is a line to the right of all $\alpha$-singularities.

We can rewrite the partition function as:
\begin{equation}
{\cal Z}=
   \int \frac{d\alpha}{2\pi i} e^{\alpha N_\sigma^3} \sum_{\sigma_x}
       e^{ -\frac{1}{2} \sigma_x {\cal A}_{x-x^\prime} \sigma_{x^\prime}
         }
\label{24}
\end{equation}
where
\begin{equation}
{\cal A}_{x-x^\prime}=
\alpha\delta_x^{x^\prime} -
\sum_{n=1}^3
   \kappa_n^\prime \delta_x^{x+n}
= \int ( \alpha - \sum_{n=1}^3 \kappa_n^\prime \cos\phi_n )
      e^{i\phi(x-x^\prime)} d^3 \phi
\label{25}
\end{equation}
The correlation function $\langle\sigma_{x}\sigma_{x+R}\rangle$
can be calculated as the derivative of generation function over sources:
\begin{equation}
\langle\sigma_{x}\sigma_{x+R}\rangle=\langle\sigma_{0}\sigma_{R}\rangle=
\frac{1}{{\cal Z}}\frac{\partial}{\partial\eta_{0}}\frac{\partial}{\partial
\eta_{R}}
\int d\alpha \sum_{\{\sigma\}}e^
{\alpha N-\sigma_{x}{\cal A}_{x-x^{\prime}}\sigma_{x^{\prime}}+
\eta_{x}\sigma_{x}}
\label{26}
\end{equation}
and after shifting integration's variables
we have
\begin{eqnarray}
\langle\sigma_{x}\sigma_{x+R}\rangle &=& \frac{\partial}
{\partial q_{0}}\frac{\partial}{\partial q_{R}}\int d\alpha
e^{\frac{1}{4}\eta_{x}{\cal A}_{x-x^{\prime}}^{-1}\eta_{x^{\prime}}} \\
&=& {\cal A}_{R}^{-1}=\int\frac{e^{-i\phi R}}{\alpha_{0}-\sum_{n=1}^{3}
\kappa_{n}^\prime\cos\phi_{n}}d^3 \phi\nonumber
\label{27}
\end{eqnarray}
$\alpha_{0}$ is the sadle point which is determined by the condition
\begin{equation}
\int\frac{d^3 \phi}{\alpha_{0}-\sum_{n=1}^{3}\kappa_{n}^\prime\cos\phi_{n}}\;=
\langle\sigma_{0}^{2}\rangle=1 \label{28}
\end{equation}
At $R_{n}\rightarrow\infty\;\;\;\;\phi_{n}\rightarrow 0$ and the pole
will be determined by
\begin{equation}
\frac{1}{2}\sum_{n=1}^{3}\kappa_{n}^\prime\phi_{n}^{2}=-\alpha_{0}+
\sum_{n=1}^{3}\kappa_{n}^\prime
\label{29}
\end{equation}
Introducing "symmetrical" variables
\begin{eqnarray}
\phi_{n} &=& \frac{\zeta}{\sqrt{\kappa_{n}^\prime}} \\ \nonumber
\zeta      &=& i\sqrt{2(\alpha_{0}-\sum_{n=1}^{3}\kappa_{n}^\prime)} \nonumber
\end{eqnarray}
we obtain
\begin{equation}
\langle\sigma_{0}\sigma_{R}\rangle=e^{\zeta\sum_{n=1}^{3}\frac{r_{n}}
{\sqrt{\kappa_{n}^\prime}a_{n}}}
\label{30}
\end{equation}
where $r=R_n a_n$

So, the potential between two probe sources depends on the choice of $n$.

\vspace{1.5cm}
\begin{center}
{\it THE PHASE STRUCTURE AND LIMIT $a_{\tau,\sigma}\rightarrow 0$.}
\end{center}

We will make suggestions about the phase structure in the whole
area of coupling constants and clarify the nature of the phases
previously obtained.
In the case $\kappa_\tau\simeq 0$ the 4-dimensional system transforms into
the set of independent 3-dimensional subsystems with $t=t_j$ as mentioned
already. The probe sources (the potential between them was calculated on
dual lattice) correspond to the magnetic charges placed inside cubes of the
original lattice.

The production over
space-like cube's plaquettes can be associated with
magnetic field flux through the cube's surface
\begin{eqnarray}
\prod_{cube}\Box_\sigma &=& \exp\{const\sum_{cube}\vec{B}\cdot \vec{n}\}\nonumber \\
B_k &=& \frac{1}{2}\epsilon_{kmn}{\cal F}_{mn}
\label{31}
\end{eqnarray}
and is not equal to zero
when the probe source is placed in corresponding dual
site. In the other words, the probe source of "electric" charge in the
site of the dual lattice corresponds to the monopol ("magnetic" charge)
of the original one.

As known, if for Wilson's loop $C^\star$
in the plane $[tx]$ of 4-dimensional dual lattice
we'd fix the time $t=t_0$
then the loop will pierce the plane $[zy]$
in two points (monopol - antimonopol).
Dirak's string which ties them together lies in the slice $t=t_0$
\cite{ukawa}.

If the potential between probe sources in each slice
will increase linearly with $R$ (in the region of coupling
$\gamma_\tau<\gamma_\tau^c\equiv\gamma_c^{\frac{1}{N_\tau}}$)
then the average value of Wilson's loop
$\langle W\rangle=\prod_{t=0}^T\langle s_{x_0}s_{x_R}\rangle$
(when the slices are independent)
will decrease exponentially according to area law.
Average value of the corresponding t'Hooft's loop
$\langle t'H \rangle$ must behave in the same way in the region
$\gamma_\sigma>\gamma_\sigma^c\equiv \frac{1-\gamma_c}{1+\gamma_c}$.
\begin{equation}
\langle t'H \rangle_{original}=\prod_{t=1}^T\langle m_{x_0}, m_{x_R}\rangle
=\langle W\rangle_{dual}=\prod_{t=1}^T\langle s_{x_0}s_{x_R}\rangle\sim
e^{-\alpha TR}
\label{32}
\end{equation}

It is obvious that parameters area ($\gamma_\tau$ and $\gamma_\sigma$)
falls to four sectors depending on the behaviour of the average values
of Wilson's and t'Hooft's loops.
\vspace{0.5cm}

\begin{tabular}{|l|c|c|c|r|} \hline
I & $\gamma_\tau>\gamma_\tau^c$ & $\gamma_\sigma<\gamma_\sigma^c$ &
$\langle W\rangle\sim e^{-\alpha L_C}$ & $\langle t'H\rangle\sim e^
{-\alpha^\prime L_{C^\prime}}$\\ \hline
II & $\gamma_\tau >\gamma_\tau^c$ & $\gamma_\sigma >\gamma_\sigma^c$ &
$\langle W\rangle\sim e^{-\alpha L_C}$ & $\langle t'H\rangle\sim e^{-\lambda^
\prime\Sigma_{C^\prime}}$\\ \hline
III & $\gamma_\tau<\gamma_\tau^c$ & $\gamma_\sigma<\gamma_\sigma^c$ &
$\langle W\rangle\sim e^{-\lambda\Sigma_C}$ & $\langle t'H\rangle\sim
e^{-\alpha^\prime L_{C^\prime}}$\\ \hline
IV & $\gamma_\tau<\gamma_\tau^c$ & $\gamma_\sigma>\gamma_\sigma^c$ &
$\langle W\rangle\sim e^{-\lambda\Sigma_C}$ & $\langle t'H\rangle\sim e^
{-\lambda^\prime\Sigma_{C^\prime}}$\\ \hline
\end{tabular}
\vspace{1cm}

This picture covers all four types of possible behaviour of the averages
under consideration which were found by G.t'Hooft \cite{thooft}
from the commutation relations analysis.
It seems impossible to "see" all four phases on a lattice with fixed
asymmetry ($\kappa_\tau=const\kappa_\sigma$) including the symmetrical one.
Our results are in good agreement with \cite{ukawa}
in the areas they studied on the symmetrical lattice
(line $\gamma_\tau=\gamma_\sigma$)

\vspace{1cm}
Up to now we treated the couplings $\gamma_\sigma(\gamma_\tau)$ freely enough
considering them independent. However, it is obvious that underlying constants
$g_\sigma(g_\tau)$ should depend on the lattice spacings through
the renormalization group relations. This connection may make some areas
of the square ($0\leq\gamma_\sigma\leq 1;\;\;0\leq\gamma_\tau\leq 1$)
inaccessible. To find exact borders of the accessible area we should
build the renormalization group relations on an asymmetrical lattice and
put $g_\sigma^2(g_\tau^2)$ into $\gamma_\sigma(\gamma_\tau)$. In "naive"
limit $g_\sigma^2\simeq g_\tau^2$. Introducing $\xi=\frac{a_\sigma}{a_\tau}$
we get $\frac{\kappa_\tau}{\kappa_\sigma}=\xi^2$.

Investigating the phase structure of our theory we were dealing with
effective couplings $\kappa_{\tau,\sigma}$.
Now we are interested in clarifying
whether critical values $\kappa_{\tau,\sigma}$
are within accessible area of temperatures
$\beta^{-1}=(a_\tau N_\tau)^{-1}$
(i.e., $\beta\neq 0$ and $\beta\neq\infty$) or not.
We should note that in the limit ($a_{\sigma,\tau}\rightarrow 0$)
$g_\sigma^2\simeq g_\tau^2\simeq g^2 << 1$ \cite{hasenf}.
So, $\kappa_\tau\kappa_\sigma=\frac{4 N^2}{g^4}>> 1$ (shaded region at
fig.6). Say, at $\kappa_\sigma\rightarrow 0;\;\;\;\kappa_\tau\rightarrow
\infty$ as $\frac{1}{g^4\kappa_\sigma}$ and $\xi=\frac{a_\sigma}{a_\tau}$ as
$\frac{1}{g^2\kappa_\sigma}$.
This narrows the accessible parameters area.
Moreover, at $N_\tau\rightarrow\infty$
the points 1 and 4 move to point ($\gamma_\tau=1;\;\gamma_\sigma=0$)
(see fig.6).

Taking into account that in this area of parameters
\begin{equation}
\tilde{\gamma_\tau}\simeq\exp\{-2 N_\tau e^{-2\kappa_\tau}\}\simeq
\exp\{-2\beta\frac{e^{-2\kappa_\tau}}{a_\tau}\}
\label{33}
\end{equation}
we may see the critical point $\gamma_\tau^c$ is accessible at finite
temperature $\beta^{-1}=(a_\tau N_\tau)^{-1}$  when
\begin{equation}
g_\tau^2=\frac{2 N_c\xi}{\ln\frac{1}{a_\tau\Lambda_\tau}}
\label{34}
\end{equation}
where $\Lambda_\tau=\frac{e^{-2\kappa_\tau}}{a_\tau}$.

As $\kappa_{\sigma,\tau}$ depend not only on $a_\tau$ but also on
$a_\sigma$ (directly and via $g_{\sigma,\tau}$ too), finding functional relations
between $\kappa_{\sigma,\tau}$ and $\beta$ seems impossible if we don't fix
the connection between $a_\sigma,a_\tau$ and $N_\tau$
\begin{equation}
\kappa_\tau=\frac{2N}{g_\tau^2}\frac{a_\sigma N_\tau}{\beta}\;\;\;
\kappa_\sigma=\frac{2N}{g_\sigma^2}\frac{\beta}{a_\sigma N_\tau}
\label{35}
\end{equation}
In the "naive" limit $g_\tau^2\simeq g_\sigma^2$ and
\begin{equation}
\frac{\kappa_\tau}{\kappa_\sigma}\simeq \frac{(a_\sigma N_\tau)^2}{\beta^2}
\label{36}
\end{equation}
The investigation of the $SU(2)$ gluodynamics phase structure on
asymmetrical lattice in the plane $\beta\times g^2$ is beyond the frames
of this paper and  will be done later.

The parameter $\xi=\sqrt{\frac{\kappa_\tau}{\kappa_\sigma}}$
usually is chosen arbitrarily
( $\xi_{Hamilt}=\infty$ and $\xi_{Eucl}=1$ ).
So, if this parameter is not restricted with additional condition
then changing it arbitrarily we may
reach any point of the curve $\kappa_\sigma\kappa_\tau=const$
at any small $g^2$, thereby crossing at least one line  between phases
($II$ and $IV$). This says that thermodynamical quantities depend on $\xi$
and, moreover, the jump on this parameter is possible for some of
them. It is commonly believed that changing the parameter $\xi$ should not
result in any observable effects. The renormalization group relations
make possible excluding the dependence of observed quantities on $\xi$ in
"naive" limit, thereby making lattice regularization with different $\xi$
equivalent \cite{hasenf,karsch,cas}. To save the
independence of observed quantities from $\xi$ in "naive" limit, we chose
$\kappa_\tau,\kappa_\sigma$ in precisely the same way as in
\cite{hasenf,karsch,cas}.
In the approximation $SU(N)\simeq Z(N)$
we failed  to create the renormalization group relations
with $\Lambda$ parameter being a function of $\xi$
which remove the lattice asymmetry parameter from the thermodynamical quantities
keeping at the same time "naive" limit results.
So, we suggest that
lattice gauge theories need
some additional condition which fixes $\xi$.

Although all calculations were carried out for the $Z(2)$ group,
they can be fulfilled for the $Z(3)$ group also, with small changes.
There are reasons to hope the results for $SU(2)$ gauge group
will be similar to those for the $Z(2)$ gauge group,
at least within approximations of \cite{yon}.

The authors are indebted for fruitful discussions with Prof. Adriano Di Giacomo.
They are grateful to O.A.Borisenko for the critical notes.

\Large
\pagebreak
\vfill
\begin{figure}
  \centering
  \input{fig1.pic}

  Figure 1.
\end{figure}

\pagebreak
\vfill
\begin{figure}
  \centering
  \input{fig2.pic}

  Figure 2.
\end{figure}

\pagebreak
\vfill
\begin{figure}
  \centering
  \input{fig3.pic}

  Figure 3.\\
  Points 1-4 and 2-3 -- dual symmetrical by pairs.
\end{figure}

\pagebreak
\vfill
\begin{figure}
  \centering
  \input{fig4.pic}

  Figure 4.
\end{figure}

\pagebreak
\vfill
\begin{figure}
  \centering
  \input{fig5.pic}

  Figure 5.\\
  \large
  \begin{tabular}{ll}
  $I$   & -- deconfinement of electric and magnetic charges\\
  $II$  & -- magnetic confinement, electric deconfinement\\
  $III$ & -- electric confinement, magnetic deconfinement\\
  $IV$  & -- electric confinement, magnetic confinement\\
  $V$   & -- we cannot investigate this region analytically
  \end{tabular}
  \Large
\end{figure}

\pagebreak
\vfill
\begin{figure}
  \centering
  \input{fig6.pic}

  Figure 6.
\end{figure}

\end{sloppypar}

\begin{thebibliography}{99}
\bibitem{shigem} J.~Shigemitsu, J.B.~Kogut. Nucl.Phys. {\bf B190} (1981) 365
\bibitem{hasenf} A.~Hasenfratz, P.~Hasenfratz. Nucl.Phys. {\bf B193} (1981) 210
\bibitem{creutz} M.~Creutz. Quarks, gluons and lattice. Cambridge University
Press, 1983, p.127
\bibitem{brez} E.~Brezin and J.M.~Drouffe. Nucl.Phys. {\bf B200} (1982) 93
\bibitem{pena} A.~Pena, M.~Socolovsky. DESY 83-003
\bibitem{yon} T.~Yoneya. Nucl.Phys. {\bf B144} (1978) 195
\bibitem{lusch} M.~Luscher, P.~Weisz. MPI-PhT/95-27; \\ DESY 95-056; HEP-LAT-9504006
\bibitem{savit} R.~Savit. Rev.Mod.Phys.{\bf 52}, 2 (1980) 453
\bibitem{balian} R.~Balian, J.M.~Drouffe and C.~Itzykson.
Phys.Rev. {\bf D11}, 8 (1975) 2098
\bibitem{ukawa} A.~Ukawa, P.~Windey, A.H.~Guth. Phys.Rev. {\bf D21}
(1980) 1013
\bibitem{thooft} G.'t Hooft In High Energy Physics Proceedings of the
European Physical Society International Conference. Palermo. 1975.
\bibitem{karsch} F.~Karsch. Nucl.Phys. {\bf B205} (1982) 285
\bibitem{cas} M.~Billo, M.~Caselle, A.~D'Adda, S.~Panzeri
HEP-LAT/9601020
\end{thebibliography}
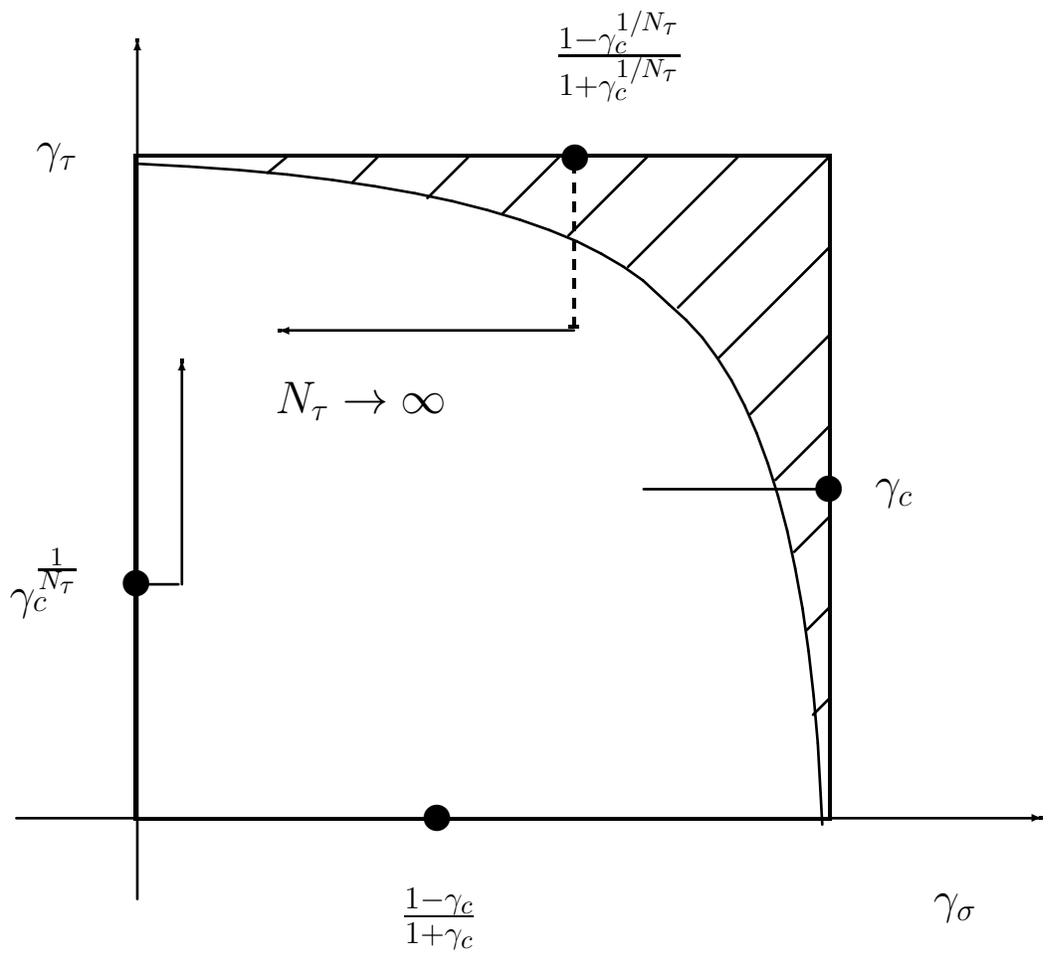
\end{document}